# Enhanced light absorption in all-polymer biomimetic photonic structures by near-zero-index organic matter

Miguel A. Castillo[1,2], C. Estévez-Varela[3], William P. Wardley[4], R. Serna[5], I. Pastoriza-Santos[3], S. Núñez-Sánchez[3], Martin Lopez-Garcia*[1],

[1]*Natural and Artificial Photonic Structures Group. International Iberian Nanotechnology Laboratory, Braga, Portugal*

[2] *Faculty of Physics/Faculty of Optics and Optometry, Campus Vida s/n, University of Santiago de Compostela, E-15782 Santiago de Compostela, Galicia, Spain*

[3]*CINBIO, Universidade de Vigo, Departamento Química Fisica, 36310 Vigo, Spain*

[4]*University of Exeter, Physics and Astronomy Department, Exeter, United Kingdom*

[5]*Instituto de Optica, Consejo Superior de Investigaciones Científicas, Madrid, Spain*

*Correspondence to: martin.lopez@inl.int

**Abstract**

Natural photosynthetic photonic nanostructures can show sophisticated light matter-interactions including enhanced light absorption by slow light even for highly pigmented systems. Beyond fundamental biology aspects these natural nanostructures are very attractive as blueprints for advanced photonic devices. But the soft-matter biomimetic implementations of such nanostructures is challenging due to the low refractive index contrast of most organic photonic structures. Excitonic organic material with near zero index (NZI) optical properties allow overcoming these bottlenecks. Here we demonstrate that the combination of NZI thin films into photonic multilayers like the ones found in nature enables broadband tuneable strong reflectance as well as slow light absorption enhancement and tailored photoluminescence properties in the full VIS spectrum. Moreover, it is shown that this complex optical response is tuneable, paving the way towards the development of active devices based on all polymer and near zero index materials photonic structures.

1. Introduction

The control over light absorption is crucial for photosynthesis, the process by which light energy is transformed into chemical energy by phototrophic organisms. However, the lack of detailed optical descriptions for such complex natural photonic systems has made it difficult to address the photonic phenomena in photosynthetic membranes[1]. Natural photonic structures provide a mechanism by which higher plants and other living organisms are able manipulate light at the nanoscale beyond the intrinsic properties of the organic compounds. It was recently discovered that some higher plants present sophisticated hierarchical photonic multilayer structures in their photosynthetic membranes (thylakoids), known as iridoplasts, that can serve to adapt light absorption capabilities to the spectral distribution available in their environment[2]. Recent developments have demonstrated that the narrowband absorption spectrum of the pigments in photosynthetic membranes is key to achieving the near unity light-harvesting quantum efficiency[3] of photosynthesis. In addition, iridoplasts present a photonic crystal with a hierarchical distribution of thylakoids, providing simultaneously and in the same photosynthetic system photonic modes and narrow absorption bands.

In natural systems the narrow absorption bands of thylakoids are provided by molecular aggregates embedded within the photosynthetic light harvesting apparatus[4]. In order to mimic the excitonic and photonic properties of photosynthetic membranes it is necessary to realize hierarchical photonic structures with similar strong narrow absorption bands. Replicating these structures with soft matter materials with similar optical properties has not yet been achieved. In this context, J-aggregates have recently attracted much attention[5] due to their narrow absorption and emission properties with properties closer to the natural counterpart in photosynthesis than oher non-aggregating dyes. It was recently demonstrated that, with high levels of aggregation, J-aggregates present narrow absorption and emission bands[6,7] that, as we show in this work, correspond to refractive indices that are close to zero Interestingly, the refractive index properties shown by thin films of J-aggregates embedded in polymeric matrices can approach values suitable for Near Zero Index materials (NZI)[8].

NZI can be achieved by materials with a Lorentzian dispersion profile[9] in which their real component of the refractive index ($n$) shows values close to zero for particular wavelengths. Their inherent strong dispersion in a narrow spectral range provides these materials with unique features such as strong phase velocity divergences or large electric field enhancements[10,11] making them very attractive in a wide range of applications, particularly as emission control and extreme non-linear interactions[12]. Most realizations of NZI materials make use of metal nanostructures showing strong optical resonances to electric field oscillations through engineered plasmonic properties[9]. These building blocks can be integrated in more complex structures such as forming an effective medium with tunable NZI properties in the visible to infra-red range[8]. Mixing NZI materials with a positive permittivity material (e.g, a dielectric) in a periodic arrangement has been used to form a photonic structure with enhanced NZI properties. This approach has recently been proposed to be capable of providing a super-absorbing close-to-darkest material in simple multilayer structures formed by ITO/TIO pairs[13]. However, NZI implementations with highly processable, low-loss or loss-free and largely tunable materials are still to be developed.

Here we propose and show experimentally that iridoplasts can work as a blueprint for the implementation of a photonic structure in which NZI properties are enhanced by the photonic multilayer design to tailor absorption to wavelengths well off-resonance of the narrow absorber maximum. We do this by doping polymer layers with J-aggregates and similarly forming a Bragg stack as the distribution of photosynthetic pigment in iridoplasts. We demonstrate that the combination of the NZI properties of the J-aggregate layers with the biomimetic photonic structuring allows for enhanced absorption at particular wavelengths. These wavelength can be selected by choosing the appropriate multilayer periodicity, allowing tunability, whilst the NZI material provides photonic bandgap splitting close to the resonance. We also demonstrate that the slow light effects[14] described for iridoplasts in the natural system can be engineered in an all-polymeric NZI system to obtain strong tunable off-resonance absorption which opens the road for the use of all organic NZI materials in several fields. Finally, we show that photoemission of the biomimetic iridoplast is tailored by the photonic dispersion bands of the heterostructure.

## 2. Results and discussion

### a. Theoretical description

Iridoplasts are natural 1D photonic crystals, also known as Distributed Bragg Reflectors (DBR), where the thylakoid membranes containing photosynthetic chromophores are evenly separated by the stroma, a homogenous non-absorbing protein-rich aqueous medium[15]. The distance between thylakoids is appropriate to create a photonic stopband in the visible range. To mimic the iridoplast we consider here a 1D photonic crystal composed of 16 unit cells, each formed by a pair of polymer thin films (i.e. 32 layers in total), as shown in Fig. 1a. One of the polymer thin films is non-doped polystyrene (PS), which will play the role of the stroma, the homogeneous layers of the natural system. PS was selected due to its ease of processing and well established optical properties as a dielectric (Fig.1b). The second constituent of the unit cell is an organic excitonic dye-doped polymer film. We selected the cyanine dye known as TDBC that can be assembled into a J-aggregate form in which the exciton de-localization makes them the most suitable candidates as biomimetic photosynthetic chromophores[16]. Moreover, since TDBC is stable in aqueous solutions, polyvinyl alcohol (PVA) is used as a polymer host-matrix.

J-aggregates are supramolecular structures with a narrow absorption band of around 10 nm in width. We can therefore consider a generic description of a material doped with J-aggregates as a highly dispersive material modelled by a Lorentz oscillator with a strong exciton transition at the central vacuum frequency $\omega_r$[7]:

$$\varepsilon_r(\omega) = \varepsilon_\infty + \frac{f_0 \omega_r^2}{\omega_r^2 - \omega^2 - i\gamma_0 \omega}, \quad (1)$$

where $\omega$ is the working frequency, $f$ is the Lorentz oscillator strength, $\gamma_0$ is the damping rate and $\varepsilon_\infty$ is the permittivity at high energy, a term that considers the dielectric constant of the material hosting the excitonic species. Therefore, for off-resonance frequencies, the dielectric constant tends towards the constant value $\varepsilon_\infty$ and it is strongly dispersive near the resonance. The optical properties of films resulting from embedding TDBC into PVA[7] are well known and can show very large oscillator strengths, as high as $f = 0.4$ which provides a strong dispersive material. Fig. 1b shows the characteristic complex refractive index for highly concentrated TDBC-PVA according to literature, where the central wavelength is $\lambda_r = 590\ nm$ ($\omega_0 = 2.1\ eV$) with an oscillator strength of $f = 0.4$ and a damping rate of $\gamma = 0.0658\ eV$[7]. We also consider $\varepsilon_\infty = 2.58$ to represent the dielectric constant of PVA. Interestingly, in the anomalous dispersion region, between 550 and 600 $nm$, the refractive index reaches very low ($n < 0.5$) and very high ($n > 3$) values below and above the excitonic resonance, respectively. Because of the very low refractive index, between $\lambda = 550$ and $585\ nm$, TDBC-PVA has a metal-like behaviour in reflectance [7] (see supplementary Fig. 1 for reflection of TDBC-PVA slab).

In this work, we study how the photonic structure modifies the absorption of TDBC-PVA films when they are structured as a DBR, similarly to the natural system[17]. The main parameter determining the photonic properties of the biomimetic iridoplast is the period of the DBR structure. The thicknesses of PS and TDBC-PVA thin film are respectively called as $d_{PS}$ and $d_{TDBC-PVA}$; therefore the period of the DBR is then $\Delta = d_{PS} + d_{TDBC-PVA}$ (see Fig. 1a). Importantly, any increase in $d_{TDBC-PVA}$ will modify the absolute absorptance of each single TDBC-PVA layer by effectively introducing more absorbing materials into the final DBR and making the comparison between structures with different periods not straightforward. For this reason, we will vary the period $\Delta$ of the structure (hence the position of the bandgap of the DBR) solely by varying the thickness $d_{PS}$ of the non-absorptive homogeneous material, while we maintain constant $d_{TDBC}$ and hence the volume of absorptive material present in the dye doped layers (TDBC-PVA).

Using a Transfer Matrix Method[18] simulation (see Methods for more detail), we calculated the reflectance at normal incidence for a fully undoped DBR (16 pairs of PVA and PS, thickness 53 nm and 90nm, respectively) and a doped DBR with the same structure but using TDBC-PVA with oscillator strength $f_0 = 0.4$ (Fig. 1c). The reflectance of the undoped structure shows a strong reflectance peak that marks the spectral position of the photonic stopband[18] given by the phase-matching condition (supplementary Note II). However, in the case of the doped DBR, two high reflectance bands are seen which can be associated with two different stopbands. We coin the terms $\lambda_+$ and $\lambda_-$ to identify the central wavelength of the stopband depending on whether it is in the spectral range below or above respectively of the anomalous dispersion region ($550 < \lambda < 600\ nm$). These regions are represented in Fig. 1C by the red and blue colours respectively. Moreover, it is also remarkable that, for the $\lambda_-$ region, the refractive index of TDBC-PVA film is below that one of the PS while, for the $\lambda_+$ region, the opposite is true. In this example, for the doped DBR, the $\lambda_-$ the stopband is far from the anomalous dispersion band and, due to the increase in refractive index contrast $\Delta n = n_{PS} - n_{TDBC}$, the absolute reflectance gets enhanced when compared to the undoped case, achieving over 80% reflectance. However, for the doped DBR in the $\lambda_+$ stopband, the Bragg reflection is close to the excitonic resonance where absorptance is high which reduces the reflectance peak through damping. By tuning the thickness of the PS layer (hence modifying period $\Delta$), it is possible to shift the $\lambda_+$ stopband of the DBR away of the anomalous dispersion band while pushing the $\lambda_-$ stopband across the excitonic wavelength $\lambda_{exc} = 590\ nm$.

Fig. 1d shows a contour plot of the reflectance at normal incidence for a DBR with 16 layers of 53 nm of TDBC-PVA while the PS thickness is varied from 60 to 200 nm. We observe a clear dependence of the reflectance on the periodicity of the structure, with an anomalous reflection region near the resonance. Interestingly, for periods with $80\ nm < d_{PS} < 180\ nm$, the photonic stopband is strongly modified by the highly anomalous dispersion. Under these conditions, the highly dispersive nature of TDBC creates a narrow band of strong refractive index contrast on both sides of the exciton band, allowing therefore simultaneously existence of $\lambda_+$ and $\lambda_-$. This phenomenon is corroborated by solving the dispersion equation (supplementary Note II) of the photonic bands for the 1D photonic crystal considered here, represented by the black line in Fig. 1c. Note that the calculations show that, for periods of $90\ nm < d_{PS} < 160\ nm$, a third stopband is obtained near resonance ($\lambda_* \approx \lambda_{exc}$). This stopband is not shown in reflectance as a result of the strong damping due to the narrowband absorption of the TDBC-PVA but it plays an important role in absorption as discussed in this work. For comparison, the dispersion relation for an undoped DBR with the same structure was plotted as a dashed red line in Fig. 1c. As expected, the regions where $\lambda_+ \gg \lambda_{exc}$ or $\lambda_- \ll \lambda_{exc}$ the stopband is tuned away from the anomalous dispersion band and therefore the photonic properties will approximate those of a DBR formed by two non-absorbing homogenous media with a refractive index contrast $\Delta n = n_{PVA} - n_{PS}$. Moreover, the second order stopband is also observed as a narrowband reflectance at low wavelengths and high $d_{PS}$.

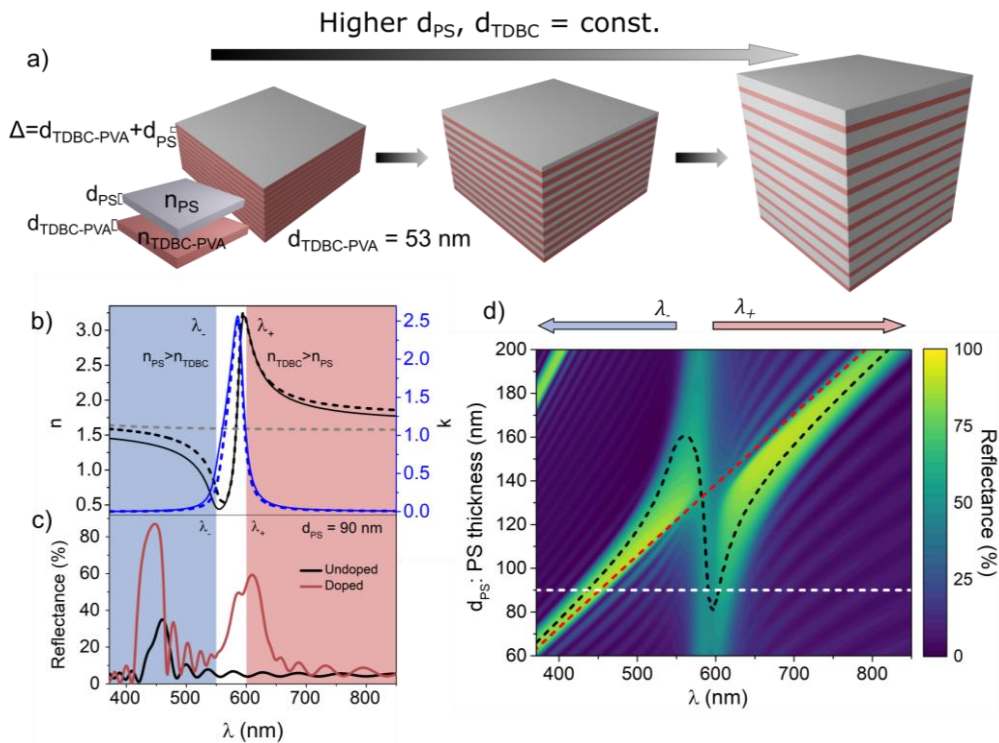

**Fig. 1 Description and optical properties of the bioinspired iridoplasts. (a)** Schematic of the Bragg mirrors composed by 16 pairs of two different materials: Polystyrene (PS) and TDBC-PVA. The TDBC-PVA thickness ($d_{TDBC}$) is kept constant at 53 nm whilst the PS thickness ($d_{PS}$) is varied. **(b)** Complex refractive index ($n + ik$) of TDBC-PVA and PS. Dashed grey line is the real

part of PS (no imaginary component) obtained by ellipsometry. Black and blue lines represent the real and imaginary component of TDBC-PVA and the solid and dashed lines represent the theoretical (using equation 1) and experimentally measured using ellipsometry refractive index (see Methods). **(c) TMM calculated** reflectance at normal incidence for undoped (black line) and doped (red line) DBR with $d_{PS} = 90\ nm$. **(d)** Normal incidence reflection of DBRs as a function of PS layer thickness. Red and black dashed lines represent photonic stopband central wavelength of un-doped and doped DBRs respectively obtained from dispersion equation solution (supplementary Note II). The white dashed horizontal line represents the cross-section plotted in sub-Figure (c) for $d_{PS} = 90$ nm.

The interactions incoming light can undergo with the biomimetic iridoplast is better analysed by calculating the electric field intensity distribution within the structure as a function of wavelength (Fig. 2a). Dispersion relation calculations of the photonic stopbands (see supplementary Note II) for $d_{PS} = 90\ nm$ show the two stopbands centred at $\lambda_- = 436\ nm$ and $\lambda_+ = 604\ nm$. It can be observed that, at normal incidence, the electric field intensity distribution at the blue edge of the $\lambda_-$ stopband ($\lambda \sim 416nm$) shows a pseudo-standing wave distribution which maximum field intensity is concentrated in the TDBC-PVA film (Fig. 2b). On the other hand, for the red-edge of the $\lambda_-$ stopband ($\lambda \sim 467nm$) the maximum field intensity concentrates at the PS thin film. This is in agreement with slow-light properties for wavelengths at photonic stopband edges[19]. The strong reduction of group velocity at those spectral ranges produces a standing wave field intensity distribution within the photonic crystal; hence increasing light-matter interactions between those wavelengths. The electric field intensity concentrates in the higher refractive index regions for the low energy band edge of the photonic stopband whilst it concentrates in the lower refractive index regions for the high energy band edge[20]. If the standing wave concentrates in the absorptive material, light absorption will be enhanced as demonstrated with broad absorbers to increase the efficiency of solar cells[14] and sensors[21]. Fig 2 demonstrates that in the case of a very narrow absorption band (i.e. strong anomalous dispersion of the refractive) the absorption enhancement by slow light could also be fulfilled for wavelengths far from resonance. These properties are studied in detail in the last section of this paper. It is also noteworthy that, for a DBR with a long enough period, the stopband shifts towards wavelengths longer than the resonance ($\lambda_+ > \lambda_{exc}$) making the refractive index contrast ($\Delta n$) change from positive to negative sign. In that case, the maximum electric field intensity concentration at the lower energy stopband ($\lambda_+$) will occur within the PS thin layer instead of the TDBC-PVA as shown in the supplementary Fig. 2. At the excitonic resonance ($\lambda_{exc} = 590$ nm), the electric field intensity distribution shows the characteristic exponential decay of an homogeneous absorbing medium.

Interestingly, it was recently proposed that natural iridoplasts might undergo self-tuning of the photonic properties to avoid photodamage. In this process, the period of the 1D photonic structures is modified to de-tune the photonic stopband from the absorption band of photosynthetic pigments[17]. The artificial iridoplast described above could be considered a simplified scenario of the natural system where the use of an NZI building block is enabling to i) experimentally tune the photonic properties, ii) study light-matter interaction enhancement at different layers of the material and iii) shift two stopbands simultaneously by simply varying the period of the system. Moreover, it is remarkable the higher refractive index contrast caused by the doping allows for a stronger reflection (~90%) than in the non-doped system (30%, Fig. 1c) providing a strong advantage for applications where large photonic strength is required. These properties open endless technological possibilities for this type of structure making it possible to obtain strong narrow band reflectance at specific wavelengths with all polymer structures and only few layers[22].

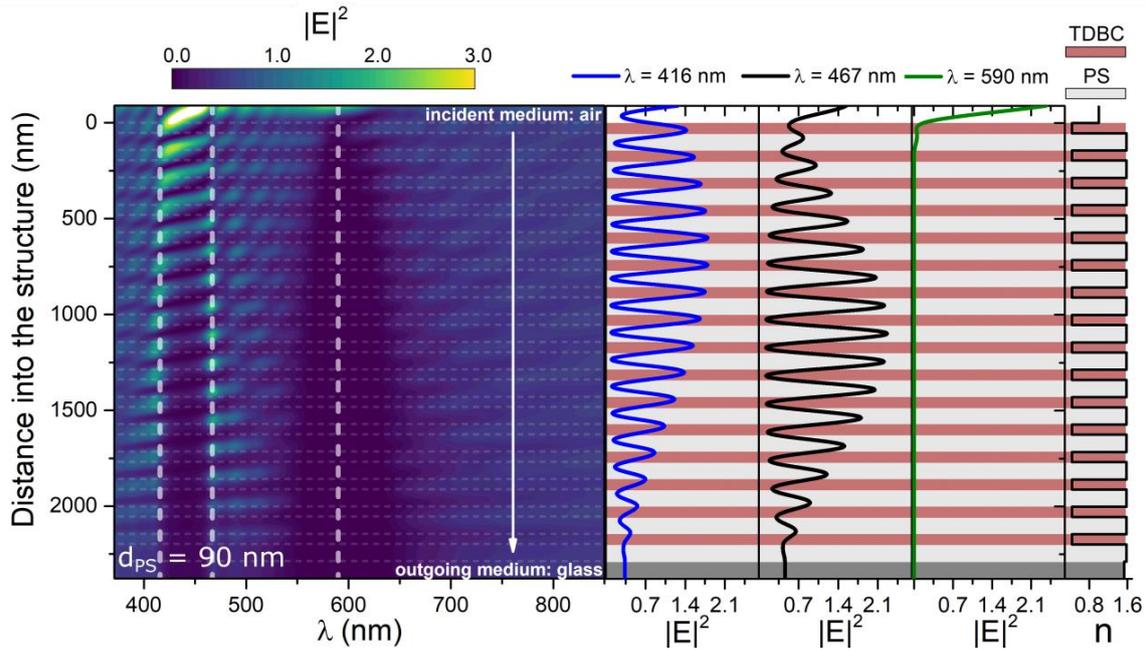

**Fig. 2 Electric field intensity inside the biomimetic structure**. **Left:** Electric field intensity ($|E|^2$) inside a TDB-PVA/PS Bragg mirror with $d_{PS} = 90\ nm$. Dashed horizontal grey lines represent the interfaces of each thin layer. Vertical dashed lines mark the wavelengths of the band-edges for the photonic stopband. **Right:** Field intensity along the vertical white dashed lines in the right contour plot. The blue line shows the high energy band-edge (416 nm), the black line shows the low energy band-edge (467 nm) and the green line shows field intensity at the exciton resonance (590 nm). Far-right axis shows the real component of the refractive index at $\lambda = 555\ nm$ for each layer (largest $\Delta n$ between TDVC-VA and PS).

## b. Experimental

To demonstrate the feasibility of the NZI biomimetic iridoplast described above, a methodology was developed for the fabrication of excitonic all-organic multilayers at different periods (see methods). Each sample is formed by 32 layers (N) with 16 of PS and 16 of TDBC doped PVA. All samples were done starting (on glass) with PS and alternating with TDBC-PVA doped layers. The thickness of the TDBC-PVA layer is maintained constant for all layers and samples at 53 nm. Therefore the period and bandgap of the DBR structure changed as a function of the PS thickness (see supplementary Fig. 3). However, as the number of layers of TDBC-PVA and their thickness is the same between samples, the total concentration of absorptive molecules is also the same between samples independently of the period. The thickness of the PS layer is varied from 88 to 207 $nm$: 88 ($\Delta_1 = 141\ nm$),

100 ($\Delta_2 = 153\ nm$), 105 ($\Delta_3 = 158\ nm$), 129 ($\Delta_4 = 182\ nm$), 147 ($\Delta_5 = 200\ nm$), and 154 nm ($\Delta_6 = 207\ nm$). The reference sample is a TDBC-PVA pseudo-slab structure ($\Delta_0 \sim 0$) formed by 16 layers of TDBC-PVA of 53 nm, each separated by just ~25 nm of PS layer, thin enough to produce negligible photonic effects outside the resonance. The structure is therefore optically equivalent to a TDBC-PVA only slab with the same total concentration of absorptive molecules that are the same that in the DBR structures.

Fig. 1b shows the TDBC-PVA and PS thin film optical properties obtained by ellipsometry (Woollam VASE Model, see methods for more information). The experimental data obtained are in outstanding agreement with the theoretically calculated ones using equation 1. While TDBC-PVA films show strong dispersive optical properties, PS films show a near-constant refractive index of value ~1.6. The refractive index of PS was also measured by ellipsometry and we always use this refractive index in the simulations. In contrast, we have used the theoretical TDBC-PVA refractive index on the previous simulations; however, from here onwards, we will always use the experimentally measured refractive indices for TDBC-PVA on all our simulations.

To analyse the absorptance of the biomimetic iridoplast, we measured reflectance ($R$) and transmission ($T$) over the same area of the samples. The absorptance at normal incidence can be calculated by $A = 1 - R - T$ assuming that scattering in other directions is negligible. To compare the absorptance quantitatively, we use the wavelength-dependent absorptance enhancement coefficient ($\gamma(\lambda)$) used in other studies[2,17]:

$$\gamma(\lambda)_{1,2,3...} = A(\lambda)_{1,2,3...}/A_0(\lambda), \quad (2)$$

where $A$ is the absorptance at normal incidence and the subscript numbers represent the samples with periods $\Delta_0, \Delta_1, \Delta_2$, etc. As a control for the absorptance of effective NZI material without photonic stopband, we consider the reference sample. The $\gamma$ parameter is a powerful tool to understand the modifications that a particular photonic environment might induce over absorptance properties of the bulk materials. In the case of the NZI biomimetic iridoplast analysed here, it allows for the comparison between the absorptance of a non-structured TDVC-PVA thin film with a DBR containing exactly the same material thickness but distributed within the photonic DBR structure. Under this definition, one could find that a given DBR can present $\gamma > 1$ (enhanced absorptance) at a particular wavelength or $\gamma < 1$ (reduced absorptance).

By the naked eye, the samples fabricated present high uniformity as shown in supplementary Fig. 3. A change in colour with the period between samples is visible not only to the naked eye but also in high magnification images (Fig. 3 and supplementary Fig. 4). Reflectance measurements at normal incidence corroborate the differences in the reflectivity colour between samples. The theoretical reflectivity corresponds well to the experiments but with a wider bandwidth for $\lambda_+$, explaining why the photonic bandgap centre in this range is slightly redder. The analysis of the reflectance over four points on the samples (see supplementary Fig. 5) showed minimal variation in the central wavelength of the photonic bandgap and absolute reflectance. This result highlights the robustness to disorder and homogeneity of the all-polymer excitonic DBRs. Moreover, as the period increases, the $\lambda_+$ stopband shifts away from the excitonic resonance. When the $\lambda_+$ stopband is outside the absorption region, the maximum absolute reflectance is achieved, with values close to 90%. In the case of the $\lambda_-$ stopband, when the period decreases it shifts away from the excitonic resonance in the region where the absorptance is low; therefore the absolute reflectance also reaches values close to 100%. However, as $\lambda_+$ and $\lambda_-$ get closer to $\lambda_{exc}$ there is a damping on the absolute reflection associated to the absorptance peak of TDBC. The DBR with an intermediate value for the DBR period, i.e. $\Delta_4$, shows the convergence of both stopbands around the exciton band with some features resembling those of a strong coupling regime.

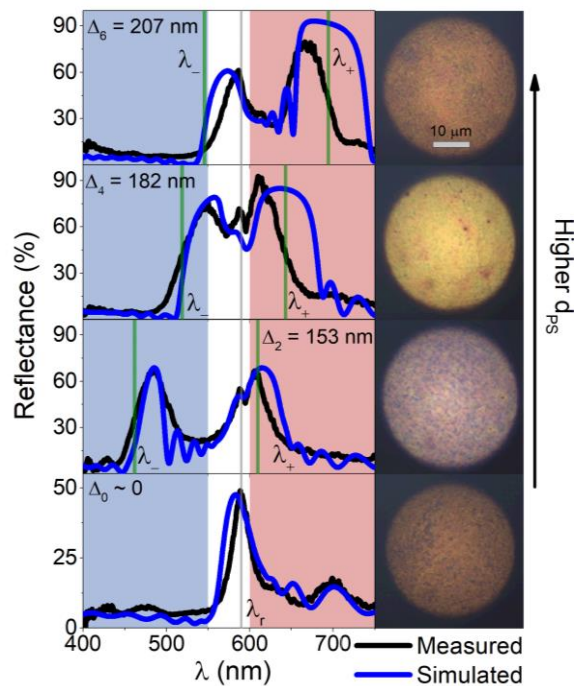

**Fig. 3 Optical characterization of biomimetic iridoplast of different periods**. **Left:** Reflection at normal incidence for samples with period $\Delta_0$, $\Delta_2$, $\Delta_4$ and $\Delta_5$ (see supplementary Fig. 4 for all samples). Black lines and blue lines represent experimental and theoretical fit, respectively. Grey vertical line marks the TDBC excitonic resonance ($\lambda_{exc} = 590\ nm$). The green lines are the central wavelengths for high ($\lambda_-$) and low ($\lambda_+$) energy stopbands calculated using the dispersion relation for the corresponding period. **Right:** Epi-illumination (reflectance) optical microscope images of the measured areas for each sample.

Next, the DBR structures were analysed by angle-resolved reflectance, measured by Fourier Image spectroscopy system coupled to a high magnification microscope that allowed us to evaluate the reflectance for incident angles $\theta < 48°$ over areas of a few µm² (see methods). Fig. 4 shows the s-polarised reflectance spectra as a function of the incident angle for three different DBRs $\Delta_2$, $\Delta_4$ and $\Delta_6$ and the reference sample (see supplementary Fig. 6 for all samples). The reflectance dependence with the incidence angle allows us to analyse the mode dispersion for both $\lambda_+$ or $\lambda_-$ stopbands in a wide spectral range for each period. As can be observed in Fig. 4, all samples with $\Delta \neq 0$ show a strong blue shift for both $\lambda_+$ and $\lambda_-$, characteristic of the dispersion curves of the photonic bandgap for a DBR[18]. For short periods, such as $\Delta_2$, the dispersion curves of the $\lambda_-$ stop band follows the dispersion behaviour of the undoped DBR but shows a higher reflectance close to 100% even with just 32 polymer layers due to the increase of the refractive index contrast. When the period of the DBR is large enough ($\Delta_{4-6}$) a third stopband appears with central wavelength $\lambda_*$. Interestingly this new stopband only appears for samples with periods where the stopband of the undoped DBR cross the excitonic resonance. For small period samples, both $\lambda_+$ and $\lambda_*$ stopbands are degenerated making them indistinguishable. However, as the period increases and the $\lambda_-$ stopband gets closer to the excitonic resonance, $\lambda_+$ and $\lambda_*$ split and the two stopbands become evident as in $\Delta_4$ (Fig. 4). When the period of the DBR is large enough ($\Delta_5$ and $\Delta_6$) $\lambda_-$ and $\lambda_*$ start merging, particularly at small angles. For $\Delta_5$, at angles $\theta > 40°$, the blueshift of $\lambda_-$ causes a split between this stopband and $\lambda_*$. Moreover, because p polarised reflection is more sensitive to angle and has a smaller full width at half maximum, this split becomes evident at lower angles (see also supplementary Fig. 6). The reflectance of $\lambda_*$ is barely visible in the experimental reflectance due to the strong narrow band absorption of the TDBC-PVA layers. However, we would like to remark that the third stopband condition is only reached in a narrow wavelength window, where large values of the refractive index of TDBC-PVA are achieved ($n \approx 3$, see Fig. 1). That wavelength restriction affects to the angular response of the dispersion curve of this third band with i) modes only supported at small angles (case $\Delta_4$) or ii) modes only supported at larger angles ($\Delta_5$ and $\Delta_6$). The experimental data is corroborated by the theoretical dispersion lines calculated numerically according to ref.[18] for each of the stopbands (black lines – Fig. 4).

We have shown that the strong dispersive optical properties of the excitonic molecular material can strongly modify the photonic crystal properties when the biomimetic iridoplast morphologies are considered. Moreover, these strong dispersive materials can achieve NZI properties acting as candidates for the fabrication of exotic absorbing materials with super-absorbing capabilities[8]. For example, Fig. 2 shows that, for a DBR with a response on the NZI wavelength region, the field can be strongly localized in the active TDBC-PVA layers at the edges of the stopband. On the other hand, it has been proposed that iridoplasts in high plants can present photosynthetic tissue structured as DBRs to induce slow light phenomena[2] and produce absorptance enhancements at wavelengths of light available in their environment. It is predicted that slow light is actively tailored within these organelles by modifying the period of the natural DBRs hence allowing for their photonic stopband to be tuned depending on the plant needs[17]. Given the optical properties of the biomimetic NZI iridoplast presented above, it is worth investigating whether we can observe similar phenomena in the artificial counterparts.

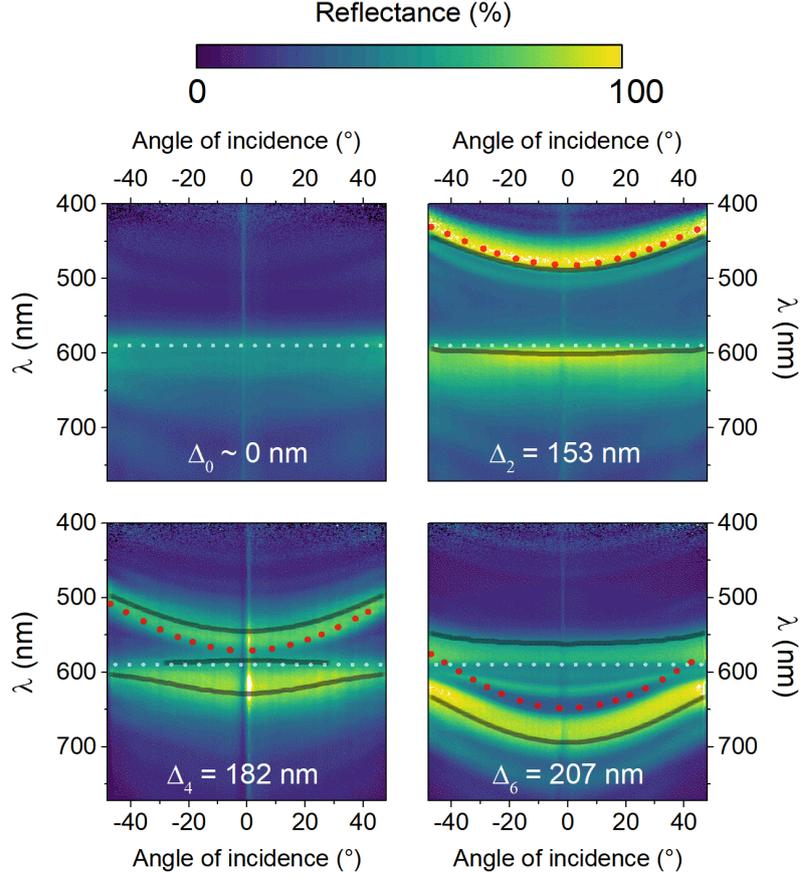

**Fig. 4 Angle resolved reflectance of biomimetic iridoplast.** Contour plots show the angle-resolved measured reflectance for the same sample in Fig. 3 (periods $\Delta_{0,2,4,6}$) under s-polarized illumination. Dashed white lines highlight the exciton band spectral position. The red dotted line shows the calculation of the dispersion of the stopband for PS-PVA undoped DBR with the corresponding period. The black solid line shows the numerical calculation of the dispersion relation of the three photonic stopbands for TDBC-PVA/PS DBRs of different periods.

Fig 5a shows theoretical and experimental data evidence for $\Delta_3 = 158\ nm$. Two well defined spectral ranges are obtained where $\gamma > 1$, demonstrating that the presence of the photonic bandgap can enhance the absorptance of the structure. In particular, the enhancement occurs at the short- and long-wavelength band edges of the $\lambda_-$ and $\lambda_+$ stopband, respectively. The enhanced absorptance is due to the slow light effect at those particular band-edges which creates an almost-standing wave with maximum field intensity at TDBC-PVA positions (as shown in Fig. 2 and supplementary Fig. 2 for the theoretical analysis) thus enhancing light absorptance, i.e. $\gamma > 1$. Interestingly $\gamma > 1$ takes place at opposite photonic stopband edges as a result of the change in the sign of the refractive index contrast ($\Delta n = n_{TDBC} - n_{PS}$) as described above.

To further explore the slow light effects over the NZI organic material, we performed reflection and absorptance analysis for the different DBR periods (represented here as PS thickness variation, $d_{PS}$).

On the blue edge of the high energy photonic bandgap ($\lambda_-$), we observe a pronounced absorptance enhancement for all $d_{PS}$, reaching values as high as $\gamma = 1.6$. In contrast, on the low energy gap ($\lambda_+$) this enhancement occurs on the red edge, reaching values of $\gamma > 2$. Absorptance enhancement ($\gamma > 1$) occurs in these areas since light is concentrated in the absorptive material. This is opposed to what happens in the red edge of $\lambda_-$ and the blue edge of $\lambda_+$ where light concentrates in the homogeneous material, reducing light absorptance ($\gamma < 1$).

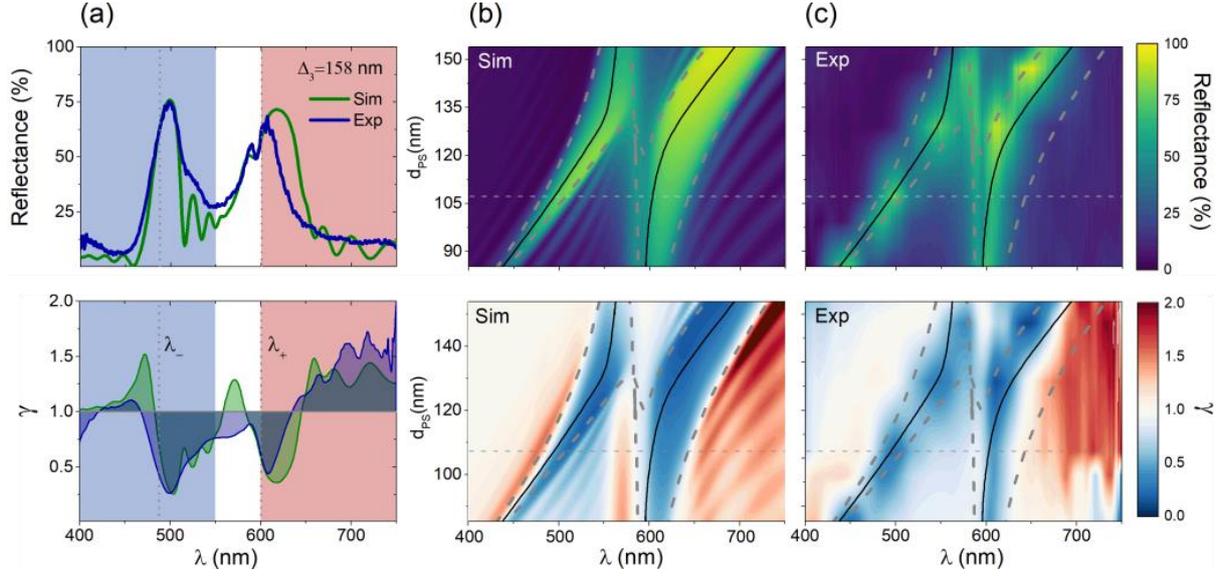

**Fig. 5 Broadband enhanced absorption by slow light in biomimetic iridoplast at normal incidence**. a) Simulated (Sim) and experimental (Exp) reflection (top) and absorptance enhancement parameter γ (bottom) for $\Delta_3 = 158\ nm$ ($d_{PS} = 107\ nm$). Dashed grey lines show central wavelength of the stopbands calculated from the dispersion relation. b) Simulated reflection (top) and γ (bottom) at different $d_{PS}$. c) Measured reflection and γ of all the excitonic DBRs fabricated. Grey horizontal dashed line highlights the spectral response for $d_{PS} = 107\ nm$ plotted in (a). The photonic band gap edges are represented by grey dashed lines. Black solid lines shows the central wavelength of the stopband calculated using the dispersion equation.

Fig. 5b shows the simulated reflectance (R) and absorptance enhancement coefficient (γ) at normal incidence for the polystyrene thicknesses ($d_{PS}$) range under consideration from the fabricated samples. Note that the reflectance plot is distinct from Fig. 1c as the refractive index used here was the one obtained from ellipsometry measurements, whilst in Fig. 1c the refractive index from equation 1 was used. The photonic band gaps observed in reflection ($\lambda_-$ and $\lambda_+$) are very clearly seen translated in the areas where $\gamma < 1$ and which redshift with an increase in period ($\Delta$). Similarly to Fig. 5a, a region with ($\gamma > 1$) is observed at $\lambda \approx \lambda_{exc} = 590$ nm and for $d_{PS} < 130\ nm$ but only for the theoretical case. This enhanced absorptance region presents almost no spectral shift upon period variation. Similarly, our calculations for the spectral position of the third stopband $\lambda_*$ (as an almost-vertical dotted black line in Fig. 1c) shows the same behaviour in this region. This therefore reinforces our hypothesis that the absorptance enhancement close to resonance is correlated to the presence of the third stopband of the DBR due to the strong anomalous dispersion of the NZI material. In addition, we have demonstrated that the closer the photonic bandgap is to the excitonic band, the higher the effect of the damping of the dipole oscillator which eventually reduces the photonic strength of the structure. Therefore, it is expected that a redshift of the stopband towards the exciton maximum absorptance will reduce $\gamma$. The opposite happens to the increased values for $\gamma$ in the long wavelength band-edge of $\lambda_+$ since, as the stopband is redshifted by the period increase, it departs from the molecular absorption band.

Despite differences in absolute values, the agreement between theoretical and experimental data is outstanding, validating the results. In $\gamma$, we again see that the photonic bandgap in the longer wavelengths ($\lambda_+$) has a wider bandwidth in the theory than in the experiments. At $\lambda_+$, the absolute values of $\gamma$ are very similar even at infra-red wavelengths ($\lambda > 700\ nm$). It is worth mentioning that the theoretical narrowband absorptance enhancement shows values of $\gamma \approx 1.2$ at $\lambda \approx 580\ nm$. This absorptance enhancement region is related to the the third stopband $\lambda_*$ not visible in reflectance due to strong narrowband absorption. Also, this spectral region corresponds to the lowest refractive index values of the TDBC-PVA, that is, is the region where the material presents a larger NZI character and therefore the interaction with the third possible stopband becomes more intense. We do not experimentally observe $\gamma > 1$ at those wavelengths. Yet, an increase in γ is observed compared to the bandgap region. The reason for this mismatch is most likely related to the fact that is not possible to produce a perfect slab with the same number of molecules as a reference. In this range, $\Delta n \approx 1.5$ which will induce enhanced absorption even for short periods like the pseudo-slab with $d_{PS} \sim 25\ nm$ used as reference in the experiments (see supplementary Fig. 7).

J-aggregate supramolecular dyes are fluorescent, therefore we can use their fluorescence to understand how light is absorbed and emitted by the photonic structures. Photoemission is intrinsically correlated to absorptance and, as a consequence, the photoluminescence of thin-film J-aggregate molecular assemblies present a narrowband emission with a small Stokes shift of a few nm to the narrow absorptance peak in most cases[5]. Interestingly, the high degrees of aggregation required to obtain NZI properties will have a dramatic effect on emission. The photoluminescence of DBRs and reference samples were measured in a Fluoro-MAX 3 spectrofluorometer (Horiba Scientific). To measure the thin films, we placed them at 45° degree to the incident light source The incident wavelength selected for the photoluminescence (PL) measurements is 550 nm for all the samples. We also performed photoluminescence excitation (PLE) where the emission wavelength is fixed at the maximum wavelength peak of the emission for each sample whilst the excitation wavelength ($\lambda_{exc}$) is scanned in the range $\lambda_{exc} = 400 - 630$ nm. By tuning the excitation source it was possible to obtain intensity and spectral measurements of the photoemission for the same sample and measurement area for excitation at different wavelengths.

Fig. 6a shows the emission spectra for the TDBC-PVA pseudo slab ($\Delta_0 \sim 0$) and the different biomimetic iridoplast fabricated under excitation at $\lambda_{exc} = 550$ nm. The emission of the TDBC-PVA pseudo slab shows an asymmetric spectrum with maximum emission at $\lambda \approx 620$ nm which is consistent with emission properties of few-molecules layers of TDBC-PVA reported in literature[23]. The slightly stronger asymmetry towards longer wavelengths is most likely due to weak Fabry-Perot optical modes supported by the thick structure. Interestingly, for increasing periods of the Bragg reflector, a continuous reduction of intensity takes place. Such intensity reduction is consistent with the position of the photonic stopbands for each period shown in Fig. 5. Beause the $\lambda_-$ stopband is redshifted at increasing periods, we will expect that samples $\Delta_{4,5,6}$ will show $\lambda_-$ stopband close to the excitation wavelength, producing a strong reflectance at $\lambda_{exc} = 550$ nm and preventing excitation illumination from reaching the inner TDBC-PVA layers of the structure which reduces the overall intensity of the emission.

At the same time, an increase in period promotes the redshift of $\lambda_+$ (Fig. 5). Fig. 6 shows the same trend with the maximum emission peak redshifting as the period increases (supplementary Note IX for set up and all samples) and the matching of the stopband with the emission bands leads to a re-shaping of the emission spectrum. This apparent red-shift is the result of the convolution of the TDBC-PVA emission with the photonic stopbands in the range $590 < \lambda < 700$ nm. As described above (Fig. 3) when the period of the DBR is large enough ($\Delta_{4-6}$) two stopbands, $\lambda_+$ and $\lambda_*$, can be found in that spectral region reducing the emission in the corresponding spectral range.

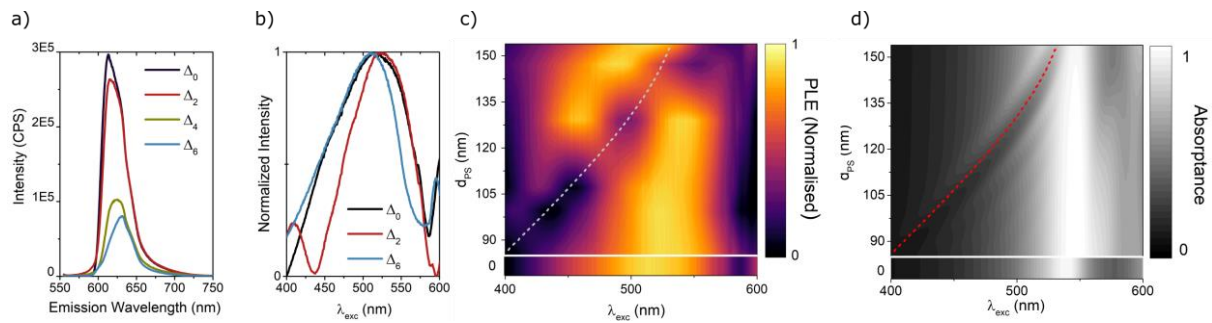

**Fig. 6 Photoemission of biomimetic iridoplast in the VIS. a)** Photoluminescence spectrum under an excitation wavelength of $\lambda_{exc} = 550nm$ and for TDBC-PVA/PS DBRs with periods $\Delta_0 \sim 0$ nm ($d_{PS} = 23$ nm), $\Delta_2 = 153nm$ ($d_{PS} = 100$ nm), $\Delta_4 = 182nm$ ($d_{PS} = 129$ nm) and $\Delta_6 = 207nm$ ($d_{PS} = 154$ nm). **b)** Normalized photoluminescence excitation (PLE) spectrum for samples with period $\Delta_0$(black line), $\Delta_2$(red line) and $\Delta_6$ (blue line). **c)** PLE measurement for periods $\Delta_{0-6}$ (see supplementary Fig. 9). White dashed line shows the position of the central wavelengths of the low energy photonic stopband ($\lambda_-$) in each case at 45°. Excitation intensity was kept constant for all samples in both PL and PLE measurements. In the case of PLE, photoemission is collected at peak intensity. **d)** Simulated absorptance of the TDBC-PVA/PS DBRs for S-polarized incident light at 45°.

It is also worth analysing the evolution of the emission maximum intensity upon excitation with different wavelengths. Fig. 6b shows photoluminescence-excitation (PLE) measurements for three samples ($\Delta_{2,4,6}$) while the contour plot in Fig. 6c shows the results for all the periods considered in this work. These spectra were obtained by fixing the collection at maximum emission for each period and varying the excitation wavelength between 400 and 600 nm. The reference ($\Delta_0 \sim 0$ nm) emission shows a uniform increase of the intensity for excitation wavelengths <520 nm while it decreases for longer excitation wavelengths, as expected from the increased reflectance of the pseudo slab due to the NZI dispersive region of the TDBC-PVA (see Fig. 1b). Overall, the biomimetic iridoplasts show a reduction of the emission intensity in agreement with Fig. 6a. It is worth noting that the periodic structures ($\Delta_{1-6}$) also show a narrowband spectral region where emission intensity is highly reduced. This reduction in emission matches the position of the photonic stopband for each period ($\Delta_2 = 473$ nm and $\Delta_6 = 575$ nm in Fig. 6b, see supplementary Fig. 9 for all periods). Fig. 6c shows that the central wavelength of the high energy stopband $\lambda_-$ at 45° follows the same pattern as the absorptance dip for an incident angle of 45° as shown in Fig. 6d. The reason for this decrease in PLE signal is the match between stopband and excitation wavelength which prevents excitation of inner TDBC-PVA molecules and hence reduces absorptance. These measurements corroborate the measurements in Fig. 5 for the modification of the absorptance by the stopbands and open the door to the study of excitonic photonic materials where absorptance and emission properties are tailored by a complex photonic environment. We observed an outstanding agreement between the calculation of the central wavelength for the stopband for each period and the drop of photoluminescence upon excitation at the same wavelength. Nevertheless, the simultaneous existence of several stopbands in the emission band of the TDBC-PVA requires further analysis that will consider not only the enhanced/prevented absorption of the excitation source demonstrated here, but also the enhancement/inhibition of emission determined by long-wavelength photonic band-edge. This analysis is out of the scope of this work and will be analysed elsewhere.

3. **Conclusions**

In conclusion, our work demonstrates that by using photosynthetic photonic systems as inspiration it is possible to tailor light-matter interactions in soft excitonic materials beyond their intrinsic properties. Our work also shows that J-aggregates are a suitable building block for NZI nanophotonics; these properties can be used to generate new phenomena in photonic structures such as multi-stopband photonic crystals or slow-light-based multi-wavelength absorptive materials with broad spectral separation. Our implementation also supports the hypothesis that the photonic structure of the natural photosynthetic photonic iridoplast uses the phenomena of slow light to enhance light absorptance at suitable wavelengths for photosynthesis. Despite the complexity of the photosynthetic natural systems, our results show the path to the demonstration of slow light phenomena in living organisms.

From a technological perspective, this work opens the door for a new set of building blocks for bioinspired NZI materials. The properties demonstrated on this work for TDBC-PVA can be expanded to the whole family of J-aggregate molecules covering the whole UV-VIS-NIR spectral range. Moreover, more elaborate nanostructuring could add further control over light-matter interaction within the structure, providing therefore a highly processable and low contaminant method for the production of NZI excitonic metamaterials.

## Methods

**Numerical calculations.** Reflectance, transmission and absorptance for all structures were obtained using an in-house implementation of the Transfer Matrix Method (TMM) [18] developed in Python programming language. Data are available upon request to M.C. Photonic bandgaps and band edge calculation was obtained from the same TMM implementation. The central wavelength of the stopband was also obtained in some cases using the phase matching condition as detailed in supplementary Note II.

**Ellipsometry measurements.** For the characterization of optical properties of the TDBC-PVA film a single layer of was deposited on a glass substrate. Spectroscopic ellipsometry characterization was carried out from the near IR-to-near UV 0.70 eV to 4.5 eV (1770 to 275 nm) using a VASE capable Woolam (WVASE) ellipsometer. The source light is a Xenon lamp, the wavelength is selected with a single monochromator, and a polarizer-retarder - sample-rotating analyzer configuration is used for the measurements (Woollam Co. Inc.). The ellipsometry data Psi ($\Psi$) and Delta ($\Delta$) data were acquired at angles of incidence (AOI) of 60º, 70º and 75º. Additionally, the transmission of the film was measured at normal incidence (AOI 0º). For the analysis, a model including the glass substrate and the film was used. The film optical constants were modelled with a Lorentz oscillator. The transmission and ellipsometric parameters $\Psi$ and $\Delta$ were analyzed simultaneously using the WVASE32 software (Woollam Co. Inc.) that allows multiangle fitting of the spectra using the transfer matrix (Abelés) formalism.

For the optical characterization of PS, a single layer of the homogeneous polymer was deposited on top of a glass substrate. The ellipsometer used was the J.A. Wollam M-2000 with a wavelength range from the UV to the near IF: 371 to 1000 nm. The source light is a 50W QTH lamp. Again, $\Psi$ and $\Delta$ data were acquired at angles of incidence (AOI) of 55°, 65° and 75°. The model to analyze the data included the glass substrate and the film used. The film optical constants were modelled using the Cauchy's equation. The results were in agreement with previously published data[24].

**Sample preparation.** The biomimetic iridoplast structures were fabricated on a coverslip (170 μm thickness) by sequential deposition of PS and TDBC-PVA layers by spin-coating. In the case of TDBC-PVA films, TDBC water solution was mixed with PVA by using a 3:1 mixture of 6.0% wt poly(vinyl alcohol) (PVA: Aldrich PVA Mw = 85000–124000) and 2% wt of J-aggregate molecules (TDBC: 5,6-dichloro-2-[[5,6-dichloro-1-ethyl-3-(4-sulphobutyl)-benzimida-zol-2-ylidene]-propenyl]-1-ethyl-3-(4-sulphobutyl)-benzimida-zolium hydroxide, sodium salt, inner salt) both in water following an already published protocol[7]. Further stirring of a mix of both solutions was performed. This solution was then diluted further with water in a 3:1 ratio (3 parts solution and 1 part water). For the deposition of PS thin films a solution of polystyrene (Aldrich, $Mw \sim 192\,000$) diluted 2.9% wt in toluene was used. The sequential deposition of polar solvent on a polymer diluted on a non-polar solvent (or vice versa) ensured minimal damage to the underlying layer, thus enabling process ability and flexibility to the fabrication[25]. The thickness of the PS layer is different for each of the doped DBRs, something controlled solely by the rpm speed during the spin coating process from 2000 to 6000 rpm (supplementary fig. 3). TDBC-PVA is deposited at 4500 rpm to yield the same 53 nm thick layer of TDBC-PVA, for both doped DBRs and pseudo-slab. This makes the tuning of the photonic bandgap by PS deposition simpler since only one solution concentration is needed. To fabricate a 25 nm layer of PS for the pseudo-slab, a 1% wt PS solution at 6000 rpm was used.

**Reflectance and transmission measurements.** Samples were characterized using a Fourier image spectroscopy setup coupled to a high-magnification optical microscope[26]. No sample preparation was required for the measurements. For reflection, samples were inspected with a tungsten-halogen white light lamp covering UV-VIS-NIR spectral range. Samples were probed focusing and collecting through either a numerical aperture ($NA$) lens (Olympus Plan N 4x, $NA = 0.10$) for normal incidence measurements or a high numerical aperture (Nikon Plan Fluor 40x, $NA = 0.75$ OFN25 DIC M/N2) in case of angle dependent reflection. Under these conditions, the measurement spot diameter was reduced respectively to $360\ \mu m$ and $40\ \mu m$. Normal incidence transmission measurements were performed using a different light source (Photonic, LED-Light-Source F1). Both reflection and transmission at normal incidence were collected in Fourier plane using a fibre-coupled 2000+ Ocean Optics (Dunedin, USA) spectrometer. The angle-dependent reflection measurements were measured using a spectrograph (Princeton Instruments, Acton SpectraPro SP-2150) and a CCD camera (QImaging Retiga R6 USB3.0 Color). Each reflection measurement was normalized against the reflection of an optically thick silver mirror and each transmission measurement to a bare substrate was measured under the same conditions as the DBRs.

**Acknowledgements**
The work by M.C.A, W.P.W and M.L-G was supported by the "Towards Biomimetic Photosynthetic Photonics" project (POCI-01-0145-FEDER- 031739) co-funded by FCT and COMPETE2020.C. E.-V., S.N.-S and I.P.-S. acknowledge financial support from MCIN/ AEI/10.13039/501100011033 (Grant No. PID2019-108954RB-I00) and the Xunta de Galicia/FEDER (grant GRC ED431C 2020/09). R.S. acknowledges Grant RTI 2018-096498-B-I00 funded by MCIN/AEI/ 10.13039/501100011033 and by "ERDF a way of making Europe

**Author Contributions**
M.C., W.P.W, S.N.S and M.L.-G. conceived the work and analysed the optical data. M.C. fabricated the samples and performed the optical measurements of the samples and designed and ran the optical models. C.E.-V., I.P.-S and S.N.-S. performed and analysed the photoluminescence measurements. R.S. performed and analysed the ellipsometry data. All authors contributed to the writing.

# Supplementary Note I: TDBC slab metal-like reflection

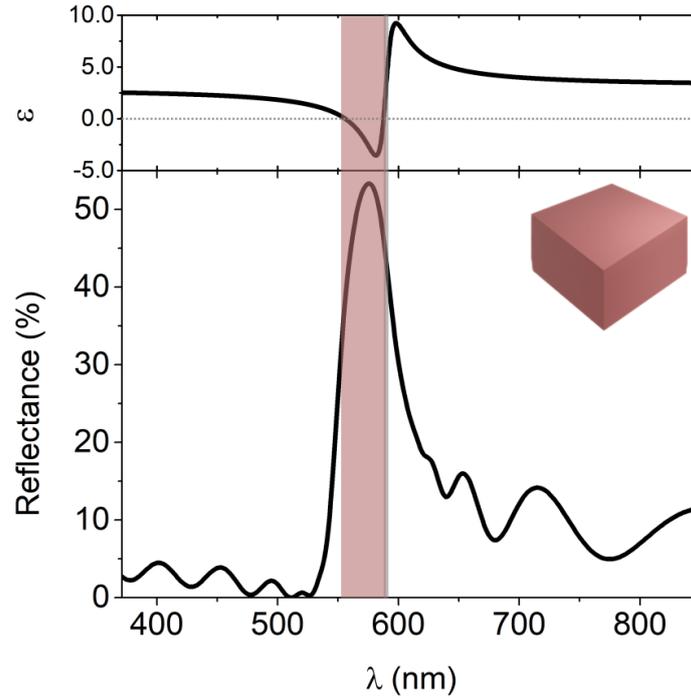

**Supplementary Figure 1.** Single slab of TDBC-PVA. **Top:** Measured real part of the dielectric constant of TDBC-PVA. **Bottom:** Reflection of a TDBC-PVA slab. For consistency, this slab has the same thickness as the total TDBC-PVA thickness in the photonic structures fabricated: $53*16 = 848\ nm$. Grey vertical line at $\lambda = 590\ nm$ highlights the exciton band resonance where absorption is maximum. Red shadowed area highlights the strong reflection range ($553 < \lambda < 590$) caused by the negative permittivity of the material which provides it with meta-like optical properties that range. Fabry-Perot interference oscillations are visible off-resonance.

## Supplementary Note II: Photonic band gap dispersion

Phase matching condition at photonic stopband fulfills:

$$k_{TDBC} \cdot d_{TDBC} + k_{PS} \cdot d_{PS} = [n_{TDBC}(\omega) \cdot d_{TDBC} + n_{PS}(\omega) \cdot d_{PS}]\frac{\omega}{c} = m \cdot \pi, \quad (2)$$

where $k_{TDBC}$ and $k_{PS}$ represent the wavevectors and $n_{TDBC}$ and $n_{PS}$ are the real part of the refractive index for the excitonic and the homogeneous medium respectively. $m$ is an integer representing the order of the stopband, which in our case is $m = 1$ since we consider only the first order.

# Supplementary Note III: Reflection and E-field intensity profile of a large period DBR

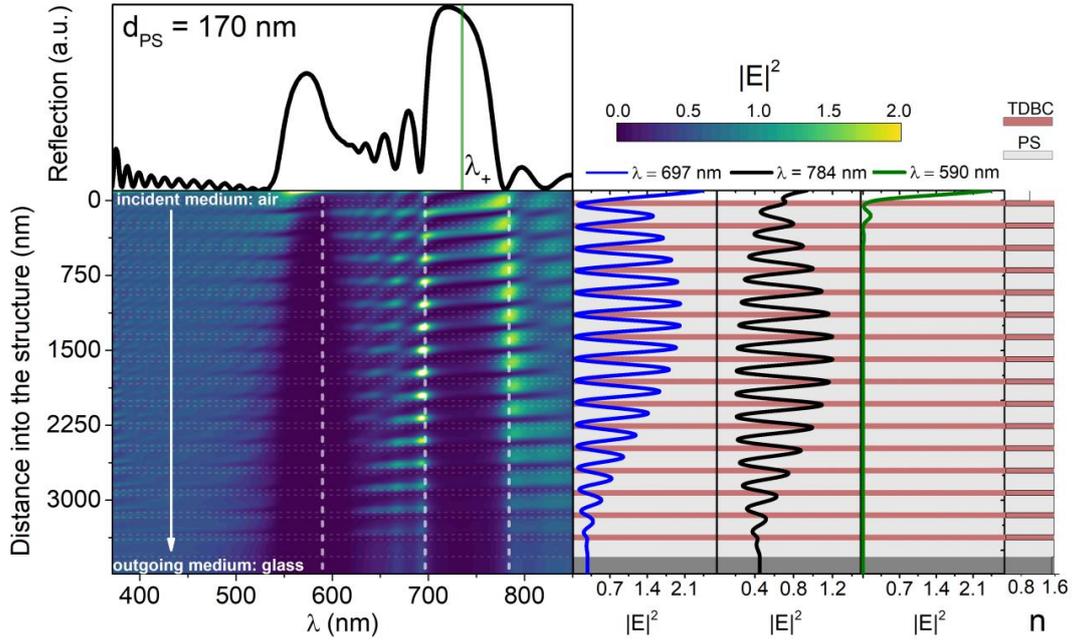

**Supplementary Figure 2:** Electric field intensity distribution inside the photonic structure. (a) Electric field intensity ($|E|^2$) inside an excitonic Bragg mirror with $d_{PS} = 170\ nm$. Field intensity along the vertical white dashed lines are represented on the right: one wavelength on each edge of the bandgap and another at the resonance. On the far right of these plots, the real component of the refractive index at λ=555 nm is represented (low for TBDC-PVA, high for PS). In contrast to figure 2 of the main manuscript, the photonic bandgap is found in the red wavelengths of the resonance, where the real part of the refractive index of PS is higher than TDBC-PVA's. For this reason, the electric field intensity maximises in the TDBC-PVA at the red edge of the photonic bandgap instead of the blue one. (b) PVA and PS (no TDBC doped layer) DBR (quarter wavelength thicknesses) centred at 590 nm. The refractive index represented here is $n_{PS}{\sim}1.58$ and $n_{PVA}{\sim}1.53$. Top plots represent the reflection of the DBRs to compare the photonic band edges to the field intensity. Dashed horizontal grey lines represent the interfaces of the layers of the systems.

# Supplementary Note IV: Spin coating calibration and sample photographs

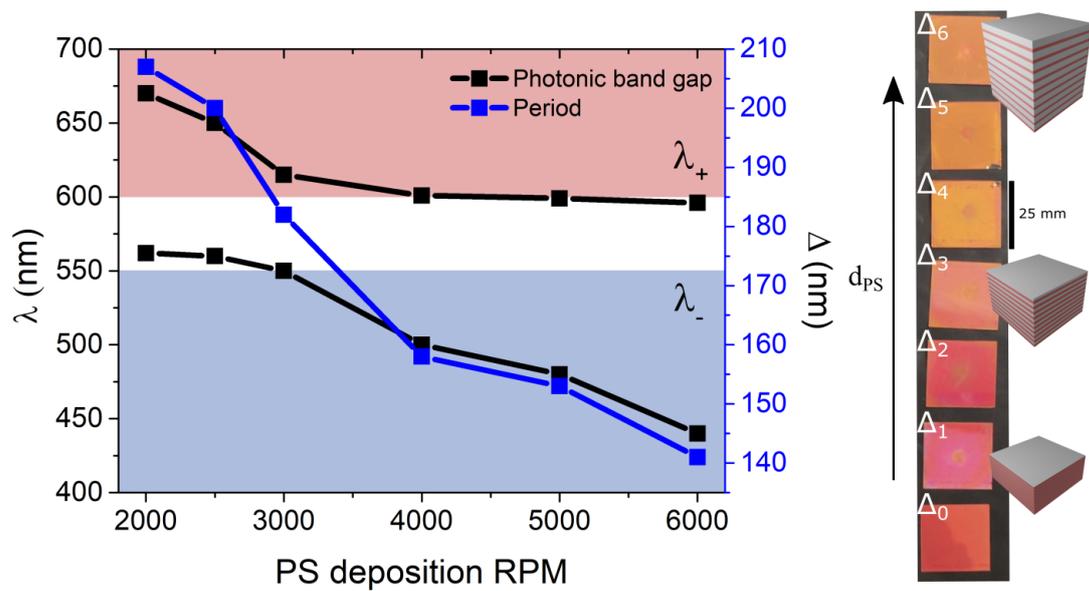

**Supplementary Figure 3:** (a) Tuning of the central wavelength of the photonic bandgap (left axis, can be either $\lambda_-$ or $\lambda_+$) and periodicity (right) by varying the rpm of the PS solution. Each sample has two photonic band gaps, however, one of them is not visible in reflectance due to the high damping of TDBC present in the white area of the graph. The only exception is the 3000 RPMs sample where both photonic band gaps are visible at normal incidence reflection. The TDBC-PVA was dropped at a constant 4500 RPM yielding an estimated thickness of 53nm. (b) Set of multilayers fabricated on $25x25\ mm^2$ coverslips. The bottom sample is a pseudo-slab of $d_{TDBC} = 53\ nm$ and $d_{PS} \sim 25 nm$ ($\Delta_0 \sim 0\ nm$) hence holding negligible photonic properties. Moving upwards, the $d_{PS}$ increases as $d_{PS} = 88$ ($\Delta_1 = 141\ nm$), 100 ($\Delta_2 = 153\ nm$), 105 ($\Delta_3 = 158\ nm$), 129 ($\Delta_4 = 182\ nm$), 147 ($\Delta_5 = 200\ nm$), and 154 nm ($\Delta_6 = 207\ nm$).

# Supplementary Note V: Normal incidence reflectance of all samples

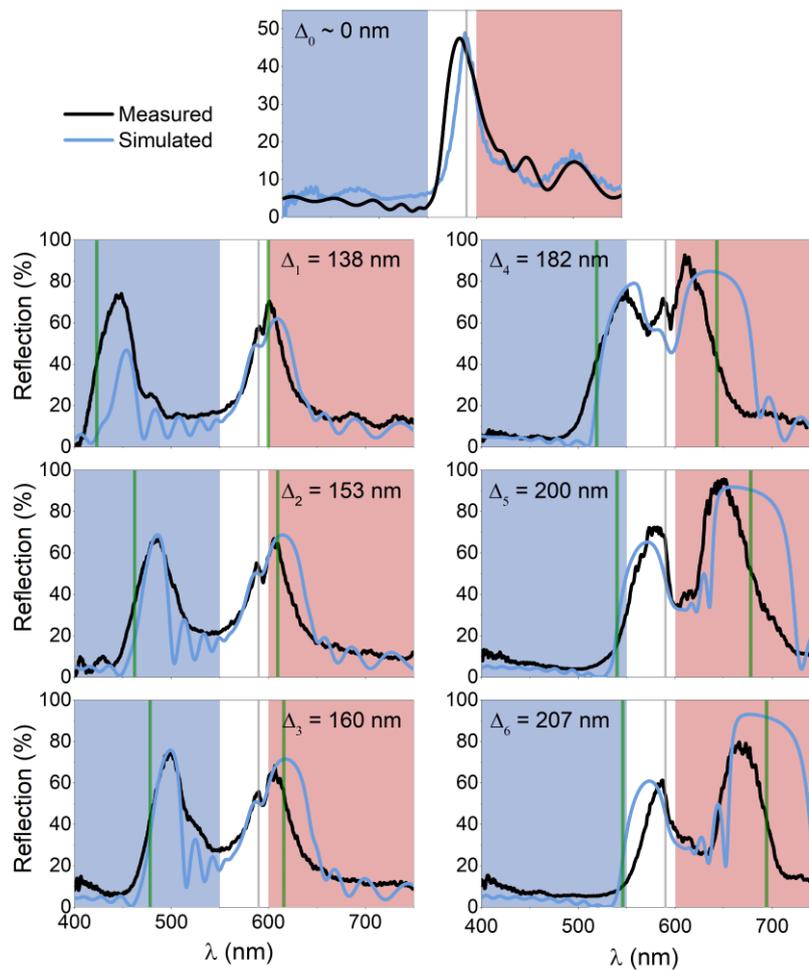

**Supplementary Figure 4:** Normal incidence reflection of all fabricated samples and the best fit simulations. Grey vertical lines represent the excitonic resonance whilst the green ones represent the solution of the phase-matching condition (supplementary section 2) and not the numerical ones.

# Supplementary Note VI: Different points on the sample

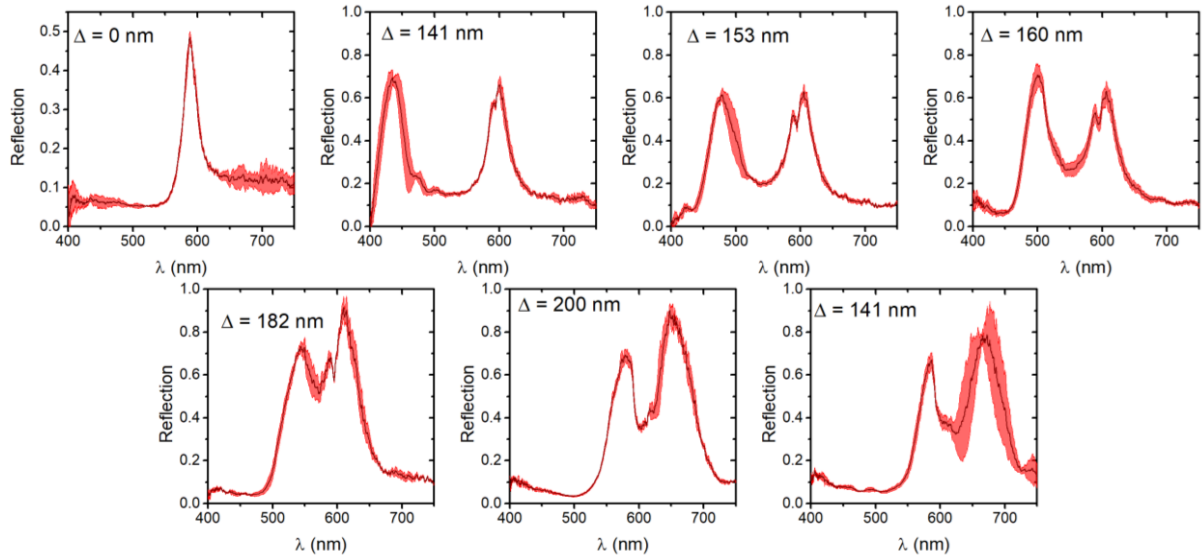

**Supplementary Figure 5: Average reflectance (absolute value) and standard deviation (shadowed red area) error for samples of different periods.** Reflectance was collected at 4 points in each sample.

# Supplementary Note VII: Angle-resolved Reflectance measurements for all samples

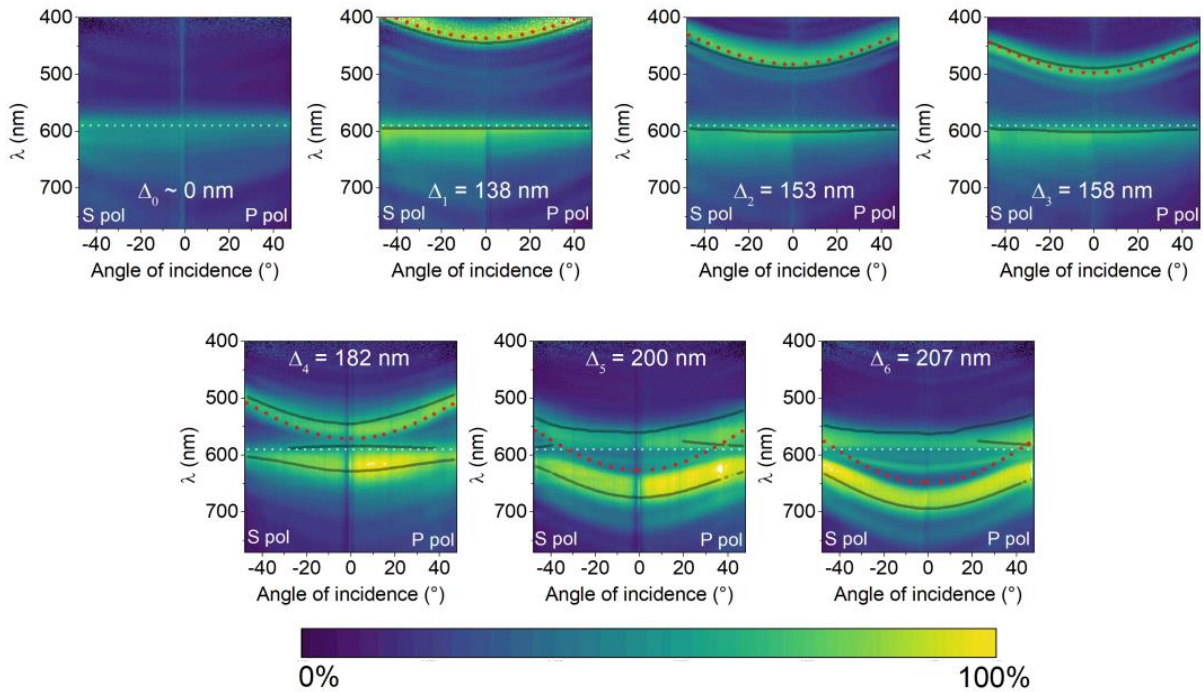

**Supplementary Figure 6: Angular reflection measurements for P and S polarized light on all samples.** Black lines represent the numerical solutions of the photonic bandgap centre. White dotted lines represent the exciton band resonance central wavelength ($\lambda = 590\ nm$) and red dotted lines the dispersion solution for undoped PVA-PS DBRs with the same period as the measured TDVC-PVA/PS structures.

# Supplementary Note VIII: Slow light at the photonic bandgap near resonance

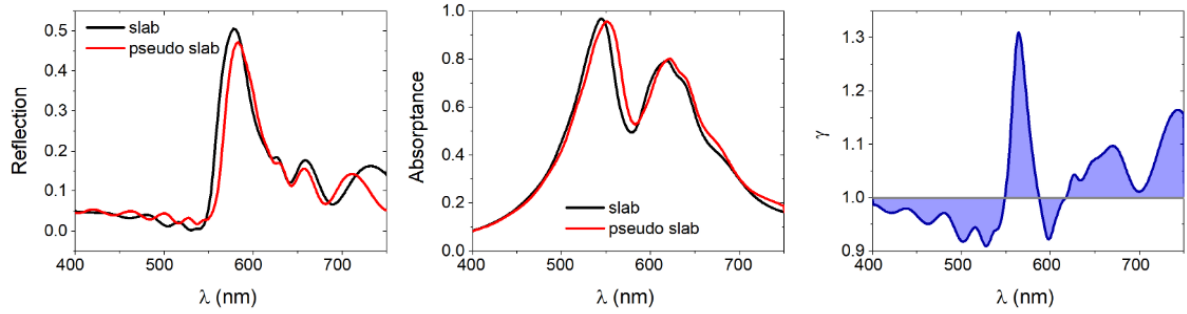

**Supplementary Figure 7: Theoretical analysis of the photonic properties of a slab and the pseudo slab.** Both have a total TDBC doped PVA thickness of $848\ nm$ ($16*53\ nm$). The pseudo slab represents the reference ($\Delta_0$) used in the experimental samples which have 16 layers of $d_{PS} = 25\ nm$. **Left:** Reflection of slab (black line) and pseudo-slab (redline). The slightly different total thickness between both structures introduces a redshift of the pseudo slab reflectance. **Center:** Absorptance of the slab (black line) and pseudo-slab (redline). Redshift is also visible in the absorptance. Absolute absorptance by the periodic structure is also enhanced in the pseudo-slab. Note that absolute absorptance is strongly reduced for both structures for the wavelengths where the samples present a negative dielectric constant and therefore a strong reflectance ($\lambda \approx 590$ nm). **Right:** Absorption enhancement factor for the pseudo slab, i.e. $\gamma = \frac{A_{pseudo\ slab}}{A_{slab}}$. Note the two regions with $\gamma > 1$. The strong absorptance enhancement at TDBC-PVA absorption peak ($\lambda \approx 590$ nm) is caused by the increase in thickness due to the presence of the PS. Even if a new pseudogap is not formed the high dispersion for TDBC-PVA near the resonance will enhance Fabry-Perot interference leading to enhancement of interaction at those wavelengths. A similar explanation applies for the region $\lambda > 620$ nm showing $\gamma > 1$. This enhancement was not observed in the DBR samples (Fig. 5b in the main text) because all absorptance spectrums were normalized by the pseudo-slab to ensure a constant number of dye molecules in all samples analysed. .

# Supplementary Note XI: Photoluminescence measurements

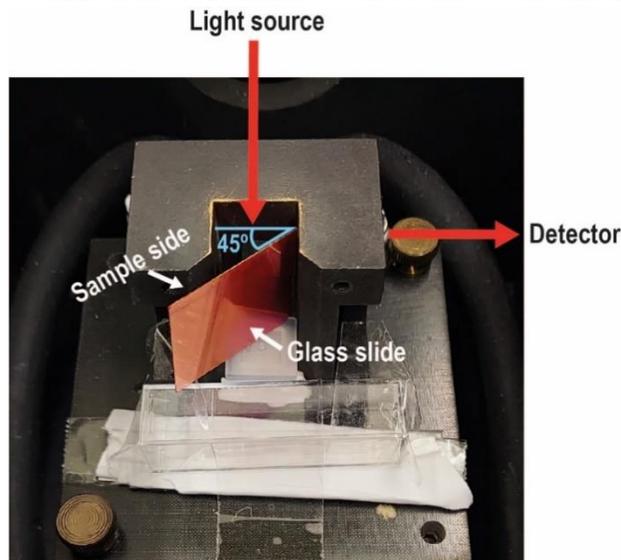

**Supplementary Figure 8: Configuration for PL and PLE characterization** of each sample in the Fluoro-MAX 3 spectrofluorometer (Horiba Scientific). Note that both excitation and collection is performed at 45 degrees to the normal of the sample.

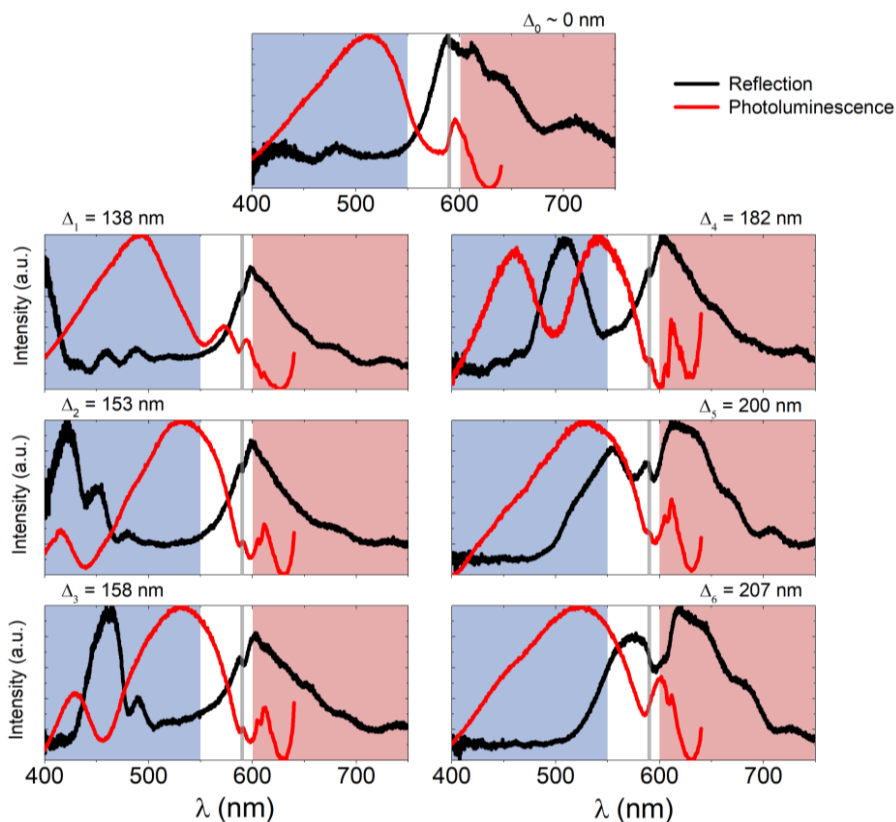

**Supplementary Figure 9:** Experimental photoluminescence (red line) and reflectance (black line) at 45 degrees for samples of different periods.